# Gate tunable supercurrent in Josephson junctions based on $Bi_2Te_3$ topological insulator thin films

Wei-Xiong Wu(吴伟雄)[1†], Yang Feng(冯洋)[1,2†], Yun-He Bai(白云鹤)[1†], Yu-Ying Jiang(姜钰莹)[1†], Zong-Wei Gao(高宗伟)[1], Yuan-Zhao Li(李远昭)[1], Jian-Li Luan(栾建立)[1], Heng-An Zhou(周恒安)[1], Wan-Jun Jiang(江万军)[1,3], Xiao Feng(冯硝)[1,3], Jin-Song Zhang(张金松)[1,3], Hao Zhang(张浩)[1,3,4], Ke He(何珂)[1,3,4*], Xu-Cun Ma(马旭村)[1,3], Qi-Kun Xue(薛其坤)[1,3,4], Ya-Yu Wang(王亚愚)[1,3*]

[1]State Key Laboratory of Low Dimensional Quantum Physics, Department of Physics, Tsinghua University, Beijing 100084, China

[2]State Key Laboratory of Surface Physics and Department of Physics, Fudan University, Shanghai 200438, China

[3]Frontier Science Center for Quantum Information, Beijing 100084, China

[4]Beijing Academy for Quantum Information Sciences, Beijing 100084, China

Supported by the Basic Science Center Project of NSFC under grant No. 51788104 and MOST of China grant No. 2017YFA0302900.

[†]These authors contributed equally to this work.

[*]Corresponding author. Email: kehe@tsinghua.edu.cn; yayuwang@tsinghua.edu.cn

Phone: 13380236780

Email: wwx13392789598@163.com; yayuwang@tsinghua.edu.cn

We report transport measurements on Josephson junctions consisting of $Bi_2Te_3$ topological insulator (TI) thin films contacted by superconducting Nb electrodes. For a device with junction length $L = 134$ nm, the critical supercurrent ($I_c$) can be modulated by an electrical gate which tunes the carrier type and density of the TI film. $I_c$ reaches a minimum when the TI is near the charge neutrality regime with the Fermi energy lying close to the Dirac point of the surface state. In the $p$-type regime the Josephson current can be well described by short ballistic junction model. In the $n$-type regime the junction is ballistic at $0.7$ K $< T < 3.8$ K while for $T < 0.7$ K the diffusive bulk modes emerge and contribute a larger $I_c$ than the ballistic model. We attribute the lack of diffusive bulk modes in the $p$-type regime to the formation of $p$-$n$ junctions. Our work provides new clues for the search of Majorana zero mode in TI-based superconducting devices.

Topological insulators (TIs) have attracted much interests due to the topologically nontrivial bulk band structure and robust Dirac-like surface states protected by the time reversal symmetry. TI-based devices have unique potentials in electrical, spintronic and quantum computation applications. It has been proposed that topological superconductivity (TSC) can be realized when the surface states of three-dimensional (3D) TIs are placed in close proximity to an $s$-wave superconductor[1]. Unlike the proposals involving semiconducting nanowires, where the spinless condition for electrons is met by tuning the chemical potential inside the small Zeeman gap induced by external magnetic field, TIs intrinsically possess spin-polarized topological surface states. Therefore, unconventional $p$-wave superconductivity and Majorana bound states may emerge in such superconductor-TI-superconductor (S-TI-S) Josephson junction[2][3][4][5][6][7], which represents a promising platform for realizing Majorana zero mode (MZM) and topological quantum computation. The S-TI-S junction has also been proposed to exhibit current-phase relationship with $4\pi$-periodicity rather than the $2\pi$-periodicity for ordinary Josephson junctions[8][9][10].

Systematic investigations of proximity-induced superconductivity on TI layers have been performed in the past few years, where Josephson supercurrent was realized in Pb and Al on $Bi_2Se_3$ single crystal[11][12], Pb on exfoliated $Bi_2Te_3$ single crystal flakes[13], Nb on $Bi_2Te_3$ single crystal[14], $(Bi,Sb)_2Se_3$ crystals[15], $Bi_{1.5}Sb_{0.5}Te_{1.7}Se_{1.3}$ crystals[16] and strained bulk HgTe[17]. Anomalous current-phase relationship and unconventional Josephson effect were observed in the weak links of exfoliated $Bi_2Se_3$ flakes[18] and strained HgTe layers[19]. Recently, the gate-tunable Josephson effect[20][21] and absence of the first Shapiro step[21] have been observed on TI thin film grown by molecular beam epitaxy (MBE), which show great potential for finding MZMs. However, to observe the MZMs, the contribution of $4\pi$ periodic current carried by ballistic SS should be larger than the $2\pi$ periodic current carried by diffusive bulk contribution[21]. Therefore, it is important to suppress the diffusive bulk contribution in S-TI-S junction, which was missing in previous studies.

Here, we fabricate Josephson junctions consisting of $Bi_2Te_3$ TI thin film grown by MBE and contacted by Niobium (Nb) superconducting electrodes. Using a gate voltage ($V_g$) to tune the carrier concentration, the Fermi energy ($E_F$) can be swept from the conduction band to the valence band. This permits investigation of the occurrence of proximity effect mediated by either electrons or holes through the charge neutrality regime with surface states alone. We find that the critical supercurrent ($I_c$) can be modulated by $V_g$, and $I_c$ reaches a minimum when the TI is near the charge neutrality regime. More interestingly, in the *p*-type regime the Josephson current can be well described by short ballistic junction model while in the *n*-type regime diffusive bulk modes emerge at $T < 0.7$ K. We attribute the lack of diffusive bulk modes in the *p*-type regime to the formation of *p-n* junctions.

The $Bi_2Te_3$ thin film is grown by MBE on $SrTiO_3$ substrate[23][24], which also acts as the dielectric material for the bottom gate (Fig. 1(a)). The thickness of $Bi_2Te_3$ thin film used here is 8 nm, which has been calibrated by RHEED. The TI film is fabricated into narrow strips with width of 500 nm by using electron beam lithography (EBL) and Ar ion etching. Josephson junctions are patterned by EBL, followed by *in situ* Ar ion etching and sputter deposition of Nb electrodes. Before sputtering the 30 nm Nb layer, a 10 nm Titanium (Ti) layer is deposited on the TI film to ensure better contact with the TI film. A series of Josephson junctions with varied distances between Nb electrodes are defined on a single chip, as shown in the optical image in Fig. 1(b). The superconducting proximity effect occurs in the two shortest junctions with $L_A =$ 134 nm (Device A) and $L_B = $ 145 nm (Device B). The supercurrent flowing through the TI thin strip is probed via conventional four-terminal ac lock-in method. As shown in Fig. 1(a), the inner two contacts are used as voltage probes while the adjacent outer two contacts are used as the current source and drain.

Figure 1(c) shows the $V_g$ dependence of sheet resistance ($R$) for Junction A at $T = 10$ K, above the critical temperature $T_c$ (Nb) = 5.9 K of the superconducting Nb electrodes. The $V_g$-dependence of $R$ provides the first indication that the device consists of thin TI films with Dirac-like topological surface states[25]. The position of the resistance maximum at $V_{CNP}^A = -68$ V indicates that there are *n*-type charge carriers in $Bi_2Te_3$, presumably due to the Te vacancies, so that extra holes are needed to locate the $E_F$ near the charge neutrality point.

Cooling the devices to below $T_c \sim 3.8$ K leads to proximity induced superconductivity in the TI film, which can be observed directly from the Josephson supercurrent[26] with zero differential resistance when dc current bias ($I$) is smaller than $I_c$. As shown in Fig. 2(a) for the measurements at $T = 20$ mK for Junction A, $dV/dI$ increases rapidly from zero to a sharp peak when the applied $I$ is larger than $I_c$. This is the characteristic behavior of the transition from superconducting state to the normal state in a Josephson junction. Both the $I_c$ value and peak $dV/dI$ value show systematic variations with the gate voltage, indicating pronounced influence of injected charge carriers on the superconducting proximity effect in TI. Figure 2(b) summarizes the same data as a color map representing the dependences of $dV/dI$ on $V_g$ and $I$. The dashed white line in this figure marks $I_c$ of the junction, which is defined by the transition of $dV/dI$ from zero to finite value. At $V_g$ = 120 V, TI is in the *n*-type region with a small resistance ~ 218 Ω at 10 K (Fig. 1(c)) and a large $I_c$ ~ 7.62 μA at 20 mK. As $V_g$ is swept from 120 V to -200 V, the $E_F$ shifts continuously from the surface conduction band ($V_g > V_{CNP}$) to the surface valence band ($V_g < V_{CNP}$). When $V_g$ reaches -60 V near the charge neutrality point $V_{CNP}^A$ = -68 V, $I_c$ reaches a minimum value of ~ 2.07 μA. The correspondence between minimum $I_c$ and $V_{CNP}^A$ indicates that the Josephson current through the junction is carried by the surface state of the TI. As $V_g$ changes to -200 V from -60 V, TI becomes *p*-type and $I_c$ increases monotonically to ~ 2.76 μA at -200 V. This characteristic $V_g$ dependence of $I_c$ demonstrates that the device operates as a bipolar Josephson junction with the supercurrent carried by electrons and holes depending on the position of $E_F$.

Figures 2(c) to (d) display the same measurements on Junction B, which shows very similar ambipolar behavior in the $V_g$ and $I$ dependences of $dV/dI$. Quantitatively, the $I_c$ values are smaller than that in Junction A, and the $dV/dI$ values are larger. The gate voltage when $I_c$ reaches a minimum is around $V_g$ = -110 V, which is also consistent with the CNP indicated by the $V_g$ dependence of resistivity. We note that $V_{CNP}^A$ is not equal to $V_{CNP}^B$ because the two junctions lie at different locations of the thin film and non-uniformity of the TI thin film can cause the fluctuation of $V_{CNP}$.

We then investigate the behavior of the junction at higher temperatures $T$ = 2.5 K and 3.8 K. With the increase of temperature, $I_c$ decreases and the peak value of $dV/dI$ at $I_c$ becomes smaller. At 2.5 K, the $dV/dI$ vs $I$ behaves like the $dV/dI$ at $T$ = 20 mK, as shown in Fig. 3(a) and (c). At $V_g$ = -70 V near the charge neutral regime, $I_c$ reaches a minimum of 1.08 μA with peak value $dV/dI$ ~ 87 Ω. These two values are smaller than that at $T$ = 20 mK. Figures 3(b) and (d) show the curves and color map of $dV/dI$ vs $I$ and $V_g$ at $T$ = 3.8 K. At $V_g$ = 120 V, $dV/dI$ still has a peak behavior, but the residual resistance ~ 13 Ω at 0.3 μA is quite big and the peak value shrinks to 28.9 Ω. Such behavior indicates that the junction loses superconducting properties at $T$ = 3.8 K at $V_g$ = 120 V. As $V_g$ decreases from 120 V, the minimum of $dV/dI$ near 0.3 μA transforms to a maximum. This $V_g$ induced transition could be explained by the Andreev reflection process that mediates the supercurrent in the Josephson junctions. When the transmission ($t$) through the barrier is low, Andreev reflection ($\propto t^2$) is suppressed compared to the normal state conductance, therefore the sub-gap conductance gets suppressed and zero-bias resistance gets enhanced. As $V_g$ decreases, the *n*-type carrier density in the TI junction region decreases as well. Once $V_g < V_{CNP}$, the carrier type in this region changes to *p*-type, while the TI regions underneath the Nb electrodes are still *n*-type due to Fermi-level pinning and screening effect of the Nb electrodes. Therefore, a *p-n* junction is formed at the edge of the junction, acting as a tunnel barrier. This barrier suppresses the Andreev reflection and thus enhances the differential resistance near zero bias. The formation of *p-n* junction barrier has been observed at the interfaces between electrodes and *p*-type TI nanoribbons[16] or *p*-type graphene[27][28].

Figure 4(a) summarizes $I_c$ vs $V_g$ for junction A at $T$ = 20 mK. There are three notable properties of $I_c$. First, $I_c$ is gate-tunable and reaches a minimum value at -60 V near the CNP, and increases monotonically on both sides. Second, the $I_c$ vs $V_g$ behavior has an opposite trend to the $R$ vs $V_g$ behavior at 10 K. As Fig. 1(c) shows, $R$ takes maximum at $V_{CNP}^A$ and $I_c$ decreases monotonically when $V_g$ tuned away from $V_{CNP}^A$. In the *n*- or *p*-type regions with larger carrier density, $I_c$ becomes larger as expected for the conventional Josephson junction model. Third, there is an apparent electron-hole asymmetry in the $V_g$ dependence of $I_c$ and $R$. In Fig. 1(c) and Fig. 4(a), $I_c$ and $R$ both change more rapidly at the *n* region than at the *p* region. We attribute this behavior to the *p-n* junction formed at the interface at $V_g < V_{CNP}^A$ as discussed above. The *p-n* junction becomes a barrier and causes $I_c$ to increase and $R$ to decrease slowly at the *n* region as $V_g$ decreases. Together with the behavior of $dV/dI$ at 3.8 K in Fig. 3(c), we believe the *p-n* junction mechanism plays a major role in the TI-based Josephson junction.

Quantitatively, the critical current reaches a minimum of 2.07 µA at $V_g$ = -60 V with critical current density $j_c = 5.18 \times 10^{-4}$ µA/nm². It is larger than TI-based Josephson junctions reported before[20][21][21], which demonstrates the high transparency at the interface between contact and TI in our device. The product $I_cR_n$ ($R_n$ is normal state resistance) provides information about the superconducting gap ($\Delta$). Figure 4(b) displays the $I_cR_n$ vs $V_g$ of junction A at $T$ = 20 mK. The normal resistance $R_n$ is taken at $I$ = 20 µA where $dV/dI$ does not vary. The $I_cR_n$ reaches minimum of 71 µV at -20 V and maximum of 230 µV at 120 V. In comparison to the induced superconducting gap $\Delta/e$ = 577 µV in the TI film with $T_c$ = 3.8 K, the $I_cR_n$ of the junction is large and this implies that the transparency of the interface is excellent. The $I_cR_n$ of our junction is comparable to that reported previously based on TI thin film grown by MBE[20][21][21].

The temperature dependence of $I_c$ is shown in Fig. 4(c) for three representative gate voltages. Near the CNP, i.e. $V_g$ = -80 V, $I_c$ increases with decreasing $T$ and saturates near the base temperature. Such behavior can be perfectly described by the ballistic short junction model 错误!未找到引用源。[29][30]. For a short ballistic junction, the temperature-dependent supercurrent can be expressed as

$$I(\varphi) = N \frac{e\pi \Delta(T)}{h} \sin(\frac{\varphi}{2}) \tanh(\frac{\Delta(T)\cos(\frac{\varphi}{2})}{2k_B T}),$$

where $N$ is the number of modes in the junction, $e$ is the electron charge, $h$ is the Planck constant, $k_B$ is the Boltzmann constant, $\varphi$ is the phase difference between two superconducting electrodes, and $\Delta(T)$ is the superconducting gap of the junction. We assume $\Delta(T)$ follows the BCS mean-field equation

$$\Delta(T) \sim \Delta_0 \tanh(1.74\sqrt{\frac{T_c}{T} - 1})$$

with $\Delta_0 = \Delta(T=0) = 1.76 k_B T_c$ with $T_c$ = 3.8 K. The critical current $I_c$ is the maximum of $I(\varphi)$ over $\varphi$. In this model, $N$ is associated with $V_g$ because the gate voltage tunes the density of Cooper pairs that carry the supercurrent modes, which corresponds to $N$ in the short ballistic model. We calculate the $I_c(T)$ for $V_g$ = 120 V, -80 V and -200 V, as shown by the solid curves in Fig. 4(c). At $V_g$ = -80 V and -200 V, the measured $I_c$ vs $T$ can be fit quite well with the calculation except for the slight decrease at $T$ < 0.5 K, which cannot be explained by this model. Therefore, the surface state and hole-doped bulk state of the TI carry supercurrent in ballistic modes. At $V_g$ = 120 V, the data match the theory of ballistic junction model at high temperature but deviate from the model at $T$ < 0.7 K, where $I_c$ is bigger than the calculated value. We attribute this deviation to the entrance of diffusive bulk modes at low temperature. The deviation part of $I_c$ can be fit well by the long diffusive junction model[31] with $I_c \propto T^{3/2}\exp(-L/L_T)$ between 0.1 K and 0.7 K. Here, $L$ is the length of the junction and $L_T = \sqrt{\hbar D/2\pi k_B T}$ is the diffusive coherence length with $D$ being the bulk diffusion constant. The calculated bulk diffusion constant $D \approx 1.79 \times 10^{-4}$ m$^2$/s is reasonable for TI and the calculated diffusive coherence length at 0.7 K

$$L_T = \sqrt{\frac{\hbar D}{2\pi k_B T}} \approx 17.6 \text{ nm}$$

is comparable to that in Ref. [21]. The diffusive bulk modes appear at lower $T \sim 0.7$ K $< T_c$ because they decay exponentially with $L/L_T$, thus becomes significant at lower $T$ when $L_T$ becomes long enough. Moreover, at $T = 0.1$ K the contribution to $I_c$ from the diffusive modes is $\sim 0.43$ μA, which is much smaller than the contribution ($\sim 5.54$ μA) from the ballistic modes. The diffusive modes are closely related to the transparency of the interface[32]. As discussed before, a *p-n* junction forms at the *p* regime and becomes a tunnel barrier. We suggest the transmission through the barrier for diffusive modes ($t_d$) is low while the transmission for ballistic modes ($t_b$) is much higher with a relation of $t_b \gg t_d$. Therefore, as Fig. 4(c) shows, in the *p* regime the junction is ballistic with no contribution from the diffusive modes. In the *n* regime the junction is ballistic at 0.7 K $< T <$ 3.8 K, while for $T <$ 0.7 K the diffusive bulk modes emerge and contribute a larger $I_c$ than the ballistic model.

This picture can also explain another strange feature observed in Fig. 2(a). In Junction A, the transition at $I_c$ is very sharp in the *p*-type regime while in the *n*-type region kink features appear in the transition process. Similar kinks were also observed in Ref. [13], which were attributed to transitions of different supercurrent components with different $I_c$. The phenomenon is consistent with our picture that in the *p*-type regime there are only ballistic modes while in the *n*-type regime there are both ballistic modes and diffusive modes. The ballistic current modes have almost the same $I_c$ while the diffusive modes have significantly smaller $I_c$. Therefore, in the *n*-type region the kink features emerge due to the presence of diffusive modes. Because the length of Junction B is larger than that of Junction A, the supercurrent has more diffusive contribution and thus there are more kinks in Fig. 2(b).

In conclusion, we investigate proximity-induced superconductivity in Josephson junctions based on high quality TI film grown by MBE. Both the normal state resistance and the critical current of the junction can be tuned by an electrical gate. The $I_c$ reaches a minimum when the TI is near the CNP, and in the *p*-type regime the Josephson current can be well described by short ballistic junction model. In the *n*-type regime the junction is ballistic at 0.7 K $< T <$ 3.8 K while for $T <$ 0.7 K the diffusive bulk modes emerge and contribute a larger $I_c$ than the ballistic model, which can be explained by the formation of *p-n* junctions. Our work also sheds new lights on the search of MZMs in TI-based superconducting devices. The vanishing first Shapiro step is an important proof of MZMs. However, in MBE grown TI-based systems, the signature of vanishing first Shapiro step is hard to detect. It may be due to the small contribution of $4\pi$ periodic supercurrent compared to that of $2\pi$ periodic supercurrent[21]. Our experiments find a new way to suppress the $2\pi$ periodic supercurrent by tuning $V_g$ to the *p*-type regime.

**Figure Captions:**

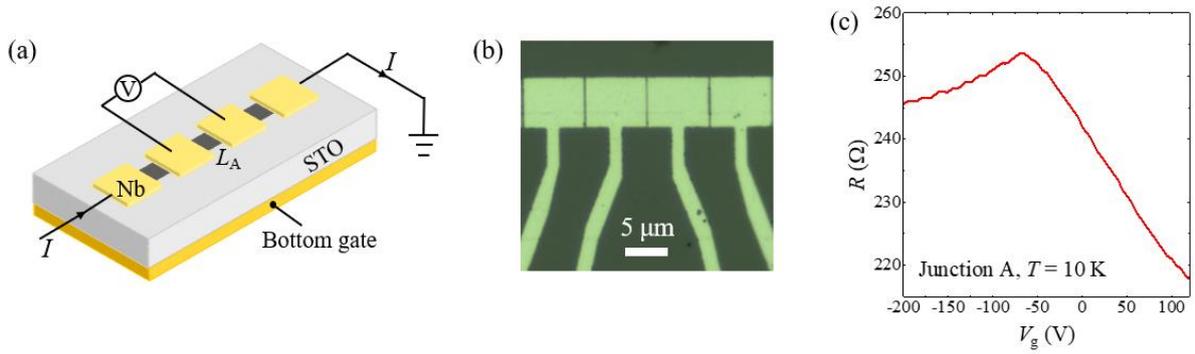

Fig. 1. (a) Schematic diagram of the S-TI-S junction device. (b) Optical image of the S-TI-S device with a scale bar of 5 μm. The junction area is defined by the two separated Nb electrodes connected by the voltmeter in the diagram. Note that the TI strip is not visible in the optical image. (c) The $V_g$ dependence of junction resistance $R$ measured at $T = 10$ K for Junction A. The maximum of resistance is located at $V_{\mathrm{CNP}}^{\mathrm{A}} = -68$ V.

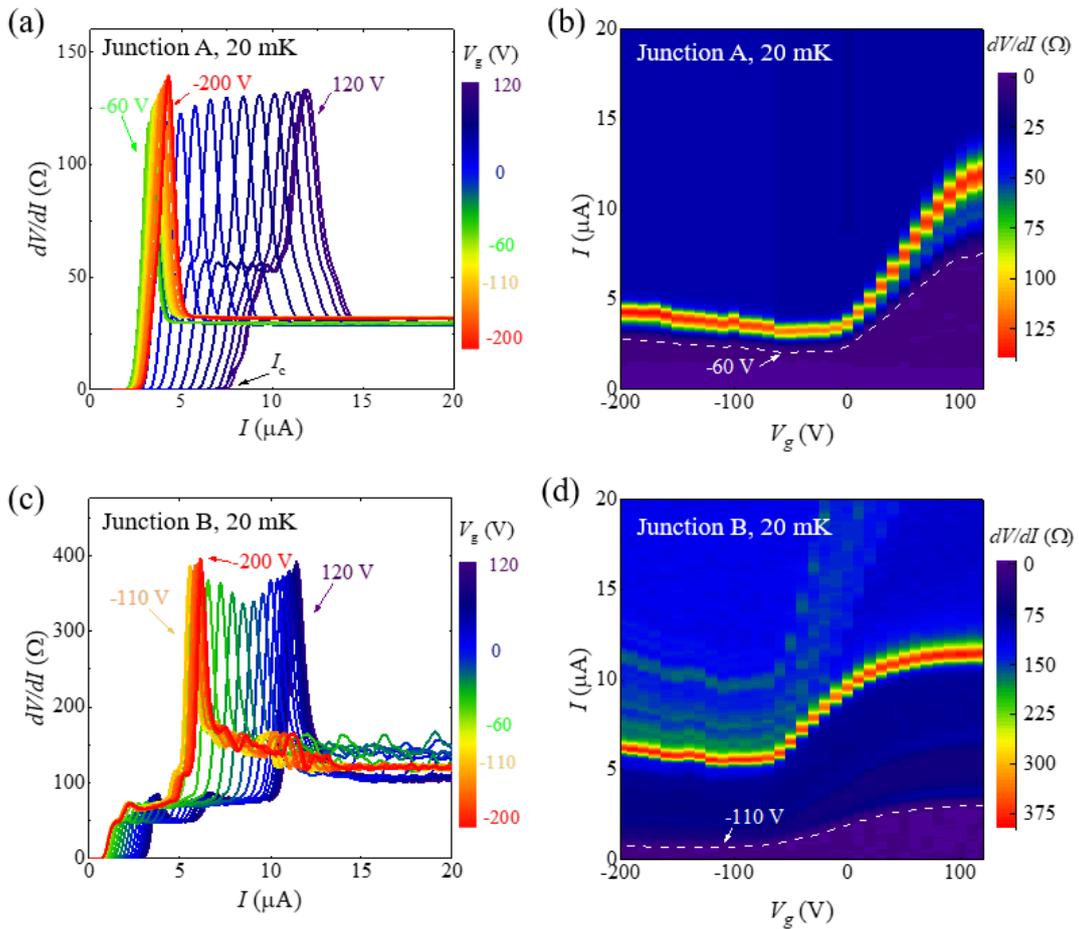

Fig. 2. Measurement of the Josephson effects for Junctions A and B. (a) The $I$ dependent $dV/dI$ curves measured at $T = 20$ mK under different $V_g$s for Junction A. (b) The color map of $V_g$ and $I$ dependences of $dV/dI$. (c) to (d) The same results on Junction B.

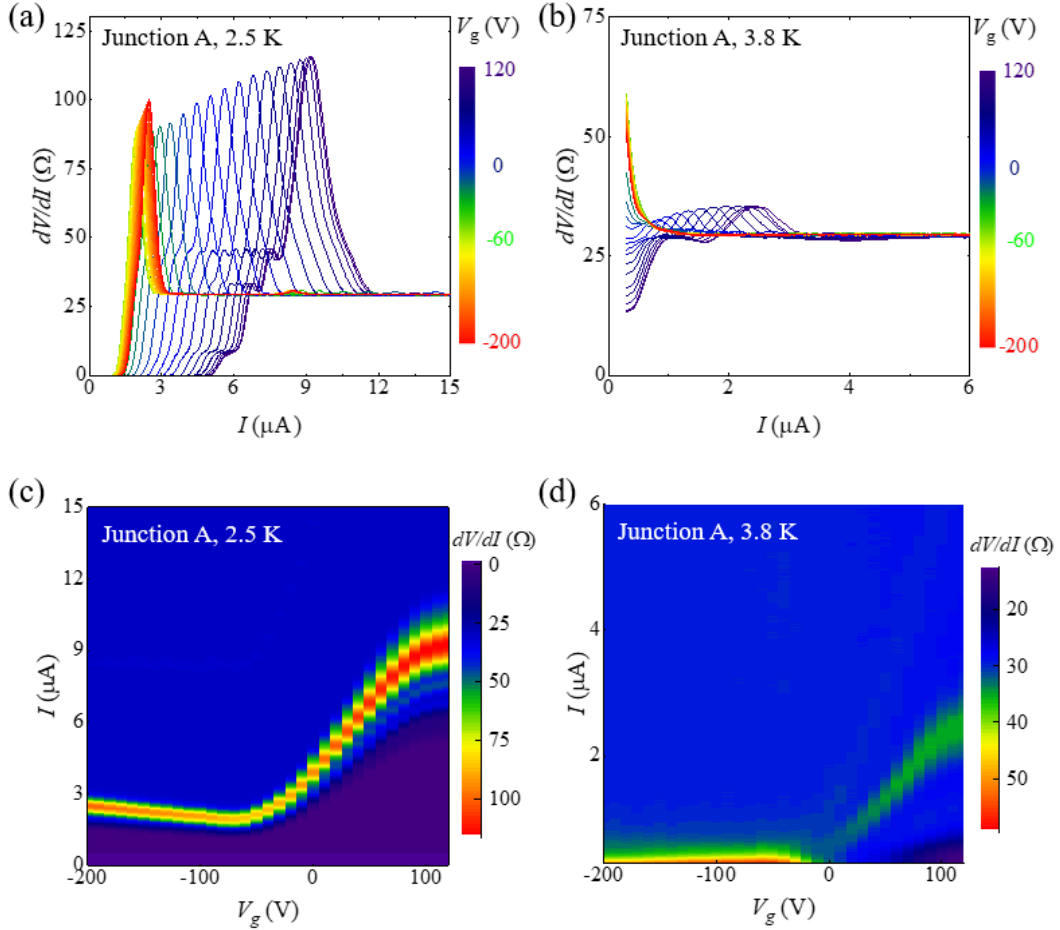

Fig. 3. Temperature dependence of the Josephson current in device A. Line cut of $dV/dI$ as a function of bias current $I$ at $T = 2.5$ K (a) and 3.8 K (b). Color map of $dV/dI$ as a function of bias current $I$ and $V_g$ at $T = 2.5$ K (c) and 3.8 K (d).

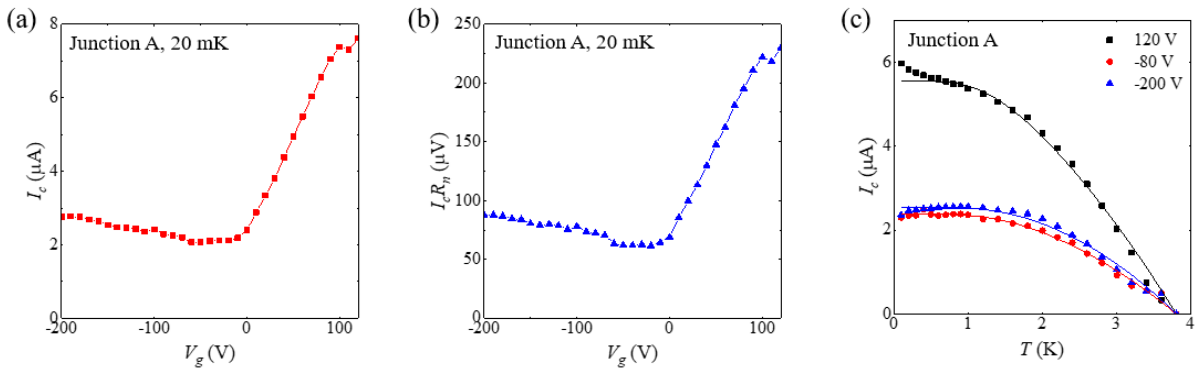

Fig. 4. (a) The $V_g$ dependence of $I_c$ for Junction A at $T$ = 20 mK. (b) $V_g$ dependence of Josephson energy $I_cR_n$ for Junction A at $T$ = 20 mK. (c) Temperature dependence of $I_c$ at $V_g$ = 120 V, -80 V and -200 V.